\documentclass[%
superscriptaddress,
 amsmath,amssymb,
 aps, physrev,
prb,
twocolumn,
]{revtex4-2}
\usepackage{xcolor} % using for editing.
\usepackage{float} % Added to fix floating?
\usepackage{physics} %added Physics package
\usepackage{subfigure} %added subfig package
\usepackage{graphicx}% Include figure files
\usepackage{dcolumn}% Align table columns on decimal point
\usepackage{bm}% bold math
\usepackage[percent]{overpic} %overpic package

\usepackage{hyperref}% add hypertext capabilities
\begin{document}

\preprint{APS/123-QED}

\title{\textbf{Fermionic versus Spin Baths in non-Interacting Transport Models} 
}% 

\author{Muhammad Zia}
\affiliation{SUPA, School of Physics and Astronomy, University of St Andrews, St Andrews, KY16 9SS, United Kingdom}
\author{Moritz Cygorek}
\email{Contact author: moritz.cygorek@tu-dortmund.de
}
\affiliation{Condensed Matter Theory, Department of Physics, TU Dortmund, Dortmund 44221, Germany}

\author{Erik M. Gauger}
\affiliation{SUPA, Institute of Photonics and Quantum Sciences, Heriot-Watt University, Edinburgh EH14 4AS, United Kingdom}

\author{Brendon W. Lovett}
\affiliation{SUPA, School of Physics and Astronomy, University of St Andrews, St Andrews, KY16 9SS, United Kingdom}


\date{\today}% It is always \today, today,

\begin{abstract}
We investigate how fermionic anticommutation shapes transport in a noninteracting resonant-level model where a single central site is coupled to an environment. To this end, we compare a fermionic reservoir with a bath of spin-$\tfrac12$ modes using exact diagonalization and a perturbative expansion of the master equation to identify the differences. Notably, we find different reduced dynamics for the spin and fermionic baths even though the particles remain noninteracting and both models enforce local Pauli blocking. These differences originate from higher-order terms in the system–bath coupling, where fermionic anticommutation introduces exchange signs in higher-order correlations. By contrast, all second-order contributions, set solely by two-point correlators, coincide. Deviations are largest for small to intermediate bath sizes, fading in the effectively Markovian regime where higher-order corrections are negligible. These results identify when spin baths can be a substitute for fermionic reservoirs and vice versa.

%\begin{description}
%\end{description}
\end{abstract}

\maketitle

%\tableofcontents
%%%%%%%%%%%%%%%%%%%%%%%%%%%%%%%%%INTRODUCTION%%%%%%%%%%%%%%%%%%%%%%%%%%%%%
\section{Introduction}
Transport through discrete quantum systems is a fundamental problem in quantum physics. Intriguing phenomena such as the Kondo effect arise in quantum transport due to many-body interactions, whose study has yielded fundamental advances in theoretical methods~\cite{Wilson1971a, Wilson1971b, Wilson1975RMP, Hewson1993Kondo, Bulla2008RMP}, and which also finds applications in experimental techniques, such as probes for the presence of local spins~\cite{Mishra2021KondoDetection}. Beyond the Kondo effect, quantum transport is central for nanojunctions \cite{Evers2020RMP,evers2018,chiechi2021,NitzanRatner2003Science,Tao2006NatNano,AgraitYeyatiRuitenbeek2003PhysRep,Reed1997Science} with possible applications as single-electron transistors \cite{FultonDolan1987PRL,Kastner1992RMP,Likharev1999ProcIEEE,GrabertDevoret1992, Pekola2013RMP_SingleElectronSources}. There, many-body Coulomb interactions are mainly responsible for reducing the active state space in a central site due to the Coulomb blockade effect \cite{Beenakker1991PRB, Aleiner2002PhysRep} while many-body correlations in the fermionic bath that describes the metallic contacts can be largely ignored. A challenge in modeling nanojunctions is the fact that real-world systems are typically coupled to additional bosonic environments, such as vibrations described by phonons, where different methods are considered depending on whether the coupling to the vibrational environment dominates over the coupling to the fermionic leads or vice versa \cite{Sowa2018BeyondMarcusI, Sowa2020BeyondMarcusII, Galperin2007VibrationalReview, Jin2008HEOMTransport, Tanimura2020HEOMPerspective, Koch2005FranckCondonBlockade, Muehlbacher2008RTPI}.

Commonly, the effects of vibrational environments are treated within the framework of open quantum systems, where the environment is traced out. For bosonic degrees of freedom, open quantum systems theory is very well established with many techniques available even conditions of strong system-environment coupling, where the environment constitutes a complex, non-Markovian memory ~\cite{breuer2002theory,rivas2012open,de2017dynamics,Gardiner2004QuantumNoise,Daley2014Trajectories,Tanimura2020HEOMPerspective,Jin2008HEOMTransport,Strathearn2018TEMPO,JorgensenPollock2019,Cygorek2022AutoCompression,Prior2010PRL_TEDOPA,Chin2010JMP_ChainMap,Makri1995JCP_QUAPI_I,Makri1995JCP_QUAPI_II,Garraway1997Pseudomodes,IlesSmith2014RC_PRA,Strasberg2018PRX_RC,McCutcheon2011VariationalPolaron,Breuer2016RMP_NonMarkov}. By contrast, a central site coupled to a bath of electrons is typically modeled using Green's functions, making use of Wick's theorem \cite{wick1950, nolting2009fundamentals}. As a step towards a unified treatment of a quantum system coupled to fermionic and bosonic environments, particularly when anharmonicities render the environments non-Gaussian and  Wick's theorem invalid, it is instructive to also consider the coupling of a central site to a fermionic bath within the framework of open quantum systems.

However, fermionic anticommutation poses subtle questions for an open quantum system formulation, e.g., when defining fermionic reduced density matrices~\cite{Coleman1963fermionDensMat}. A modern framework where the fermionic reduced density matrix is composed of two parts corresponding to even and odd particle numbers~\cite{Cirio2022} enables the formulation of fermionic influence superoperators, hierarchical equations of motion (HEOM), and generalized Lindblad master equations for fermionic open quantum systems. Numerical methods to simulate the dynamics of quantum systems coupled to fermionic baths may be derived using the Jordan-Wigner transformation, such as in fermionic TEDOPA~\cite{Nusseler2020FermionicBath} or making use of Grassmann algebra, as in the fermionic version~\cite{thoenniss2023nonequilibrium} of process tensors~\cite{JorgensenPollock2019,Cygorek2022AutoCompression}, where the Feynman-Vernon influence functional is efficiently represented using temporal tensor networks. Notably, the fermionic process tensors can also be obtained using the Jordan-Wigner mapping instead of Grassmann algebra (see Appendix B of Ref.~\cite{Cygorek2024tree}). The challenge in working with Grassmann numbers is their non-commutativity. However, the Jordan-Wigner transformation maps a fermionic system onto one composed of spin-1/2 and can thus also be used to simulate fermionic systems on gate-based quantum computers, but at the cost of introducing nonlocal string operators.

This naturally raises the question of whether it is really necessary to account for the proper fermionic anticommutation relation in quantum transport, and under which circumstances it is a good approximation to replace a fermionic environment by a bath of two-level (spin-1/2) systems while simply ignoring any possible sign change due to the fermionic anticommutation. Then, conventional open quantum systems methods for bosonic or other commuting environments could be directly transferred to systems coupled to fermionic baths. 
Noting that the most prominent phenomena arising from the fermionic anticommutation are (a) Pauli blocking and (b) the Coulomb exchange interaction responsible for effective spin-spin interactions and chemical bonding, one may expect that in the absence of many-body interactions, an effective spin bath, which also automatically enforces Pauli blocking, would lead to similar environment effects as a fermionic bath.

Here, to test this hypothesis, we consider a non-interacting transport model that is small enough to be solved by exact diagonalization for both fermionic and spin environments. Surprisingly, we find that both systems only show the same dynamics up to second order in the system-bath coupling, while for small to moderately sized baths near half filling, sizable discrepancies are identified.

%\textcolor{orange}{Here, to test this hypothesis, we focus on the minimal resonant-level model, where a single system mode couples to a structured environment—standard in transport yet simple enough for exact solutions. We contrast a fermionic bath, whose quadratic Hamiltonian is Gaussian and obeys Wick’s theorem, with a spin (two-level) bath that enforces local Pauli exclusion but omits Jordan–Wigner strings, thereby removing exchange signs between distinct sites. In a perturbative master-equation treatment, the generators coincide to second order (only two-point correlators enter) but differ at higher order, where statistics-induced exchange signs appear explicitly. Consistent with this mechanism, we find sizable discrepancies for small to moderately sized baths near half-filling, while in a scaling limit that fixes the overall Markovian rate, higher-order contributions become negligible and both descriptions converge. This delineates when spin-bath surrogates are reliable in practice and when a full fermionic treatment is required. }

The remainder of the paper is organized as follows. In Sec.~\ref{sec_theory} we define the resonant-level model in fermionic and spin formulations and summarize the perturbative master-equation approach, highlighting where algebra-dependent terms arise. Sec.~\ref{sec_results_disc} presents and discusses exact-diagonalization results across different bath sizes and compares them to the second-order master-equation prediction. We conclude in Sec.~\ref{sec_conc} by summarizing the regimes of agreement and difference and discussing implications for simulations of open quantum systems.

%%%%%%%%%%%%%%%%%%%%%%%%%%%%%%%%%MODEL%%%%%%%%%%%%%%%%%%%%%%%%%%%%%%%%%%%%
\section{Theory}
\label{sec_theory}
\subsection{Model}
We consider an open quantum system described by the total Hamiltonian:
\begin{equation}
  H = H_S + H_E,
\end{equation}
where \(H_S\) represents the system Hamiltonian, and \(H_E\) describes the environment and its interaction with the system. Explicitly, these are defined as (throughout this work, we adopt natural units, i.e., \(\hbar = 1\)): 
\begin{equation}
  H_S = \omega_0\,O_0^\dagger O_0,
\end{equation}
\begin{equation}
  H_E = \sum_{k=1}^{N_E} \Bigl[ \omega_k\,O_k^\dagger O_k + V_k\,\Bigl( O_0^\dagger O_k + O_k^\dagger O_0 \Bigr) \Bigr].
\end{equation}

\begin{figure}[H]
    \centering
    \includegraphics[width=0.8\columnwidth]{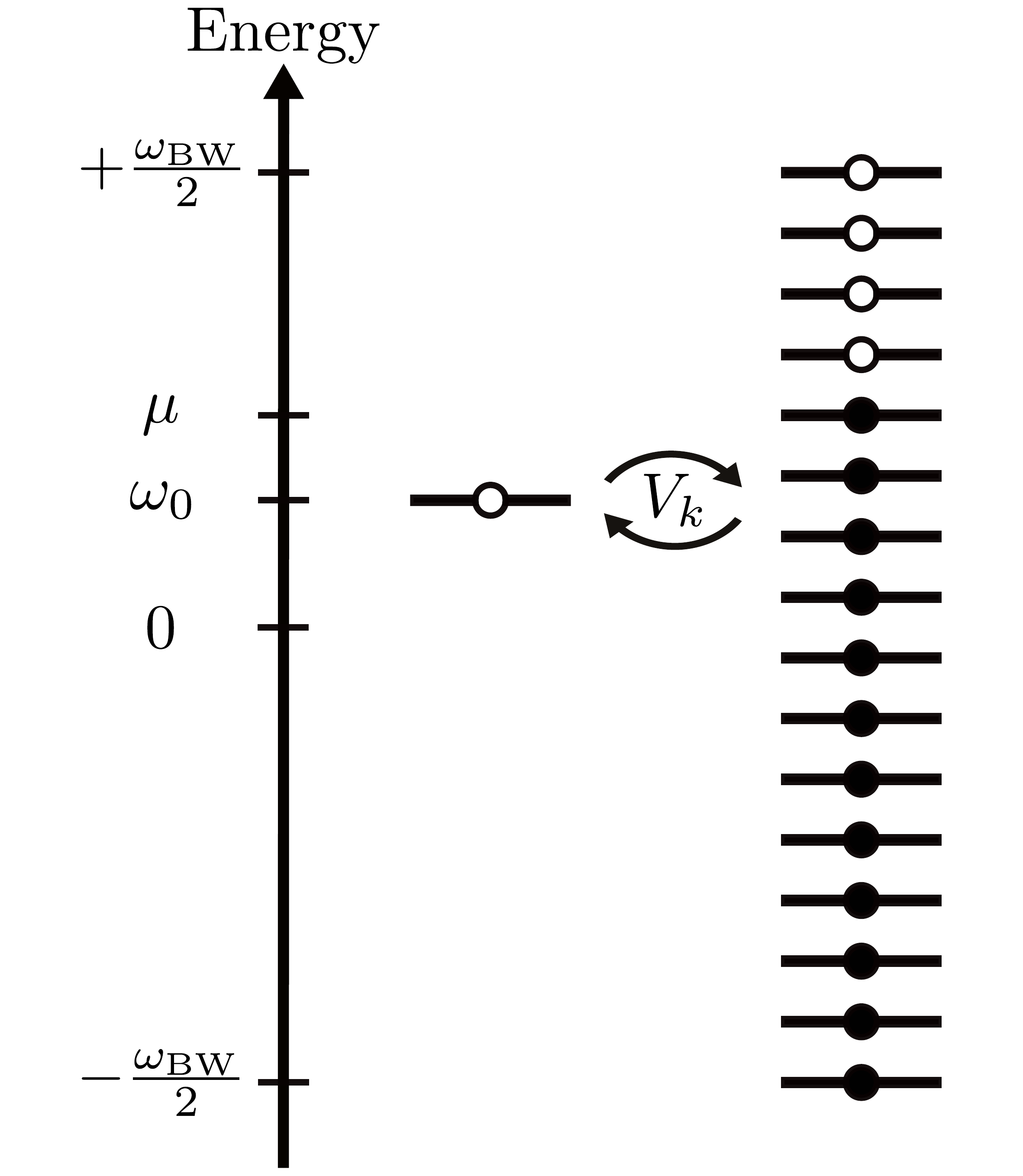}
\caption{Schematic representation of the resonant--level model: A system mode at energy $\omega_0$ is coupled (with coupling strengths $V_k$) to an environment consisting of discrete modes within a bandwidth $\omega_{\mathrm{BW}}$, partially filled up to the chemical potential $\mu$.}
    \label{fig:schematic}
\end{figure}

Here, the system is represented by a single mode at frequency \(\omega_0\), coupled to an environment composed of \(N_E\) modes with frequencies \(\omega_k\). The coupling strength between the system mode and the \(k\)--th bath mode is given by \(V_k\). The operator \(O_j\) is a generic two-level lowering operator associated with mode \(j\), and we study two distinct physical realizations:

\begin{enumerate}
    \item \textbf{Fermionic Model:}  
    Here, we set \(O_j \equiv c_j\), where \(c_j\) are fermionic annihilation operators satisfying the canonical anti-commutation relations:
    \begin{equation}
      \{c_i, c_j^\dagger\} = \delta_{ij}, \quad \{c_i, c_j\} = 0.
    \end{equation}

    \item \textbf{Spin Model:}  
    Alternatively, we choose spin (two--level) lowering operators \(O_j \equiv \sigma_j^-\). The corresponding algebra is:
    \begin{equation}
      \{\sigma_j^-, \sigma_j^+\} = 1, \quad [\sigma_i^-, \sigma_j^{\pm}] = 0 \quad \text{for } i\neq j.
    \end{equation}
\end{enumerate}
The \textit{resonant-level model} \cite{Newns1969} serves as a minimal yet powerful framework for exploring open quantum dynamics, transport, and many-body effects in non-equilibrium settings. While the fermionic and spin variants differ in their algebraic structure, both capture the essential dynamics of impurity--bath coupling. Figure~\ref{fig:schematic} illustrates the system--bath coupling common to both fermionic and spin realizations of the resonant--level model.

It is worth noting that in the fermionic model the total Hamiltonian is quadratic in creation and annihilation operators and thus admits an exact single-particle (one-body) diagonalization. By contrast, the spin model does not reduce to a quadratic form under any mode transformation, so the efficient single-particle diagonalization available for fermions does not carry over to spins. For completeness, we outlined this briefly in Appendix.~\ref{app:singlebody}.

\subsection{Perturbative Master Equations}
Master equations provide a standard yet powerful analytical framework for studying the dynamics of open quantum systems \cite{breuer2002theory}. By tracing out the environment degrees of freedom and expanding perturbatively in the system--bath coupling strengths, one obtains an effective dynamical equation governing the reduced density matrix of the system. In this section, we examine the perturbative expansion, beginning with the second-order master equation, where the dynamics remain unaffected despite the distinct operator algebra of the spin and fermionic models. At fourth order, this correspondence need not hold in general; to explore this, we analyze a representative term that naturally arises in the expansion.

\subsubsection{Second-Order Master Equation}
At second order in the coupling strength, the master equation governing the reduced density matrix \(\rho_S(t)\) can be obtained. A detailed derivation is provided in Appendix~\ref{app:secorder_equiv}. The resulting equation reads:
\begin{align}
\frac{d\rho_S(t)}{dt}
= -\int_0^t dt' \Bigl[ &C_1(\tau)\bigl[O_0,O_0^\dagger \rho_S(t')\bigr] \nonumber \\
& + C_2(\tau)\bigl[\rho_S(t')O_0^\dagger,O_0\bigr] + \mathrm{h.c.} \Bigr],
\label{eq:2nd_order_ME}
\end{align}
where, h.c. is the hermitian conjugate, and, \(\tau = t - t'\), and the bath correlation functions are
\begin{align}
C_1(\tau) 
&= \sum_k |V_k|^2 e^{-i(\omega_0 - \omega_k)\tau} \langle O_k^\dagger O_k \rangle \notag \\
&= \sum_k |V_k|^2 e^{-i(\omega_0 - \omega_k)\tau} f(\omega_k),
\end{align}
\begin{align}
C_2(\tau) 
&= \sum_k |V_k|^2 e^{-i(\omega_0 - \omega_k)\tau} \langle O_k O_k^\dagger \rangle \notag \\
&= \sum_k |V_k|^2 e^{-i(\omega_0 - \omega_k)\tau} \bigl[1 - f(\omega_k)\bigr].
\end{align}

The operators appearing in Eq.~\eqref{eq:2nd_order_ME} are kept deliberately generic and may correspond to either fermionic or spin operators. Crucially, at second order, the master equation we derive is identical for both the fermionic and spin models. This equivalence arises because the bath correlation functions depend only on occupation of bath modes, such as \(\langle O_k^\dagger O_k \rangle = f(\omega_k)\), which reflect the local Pauli--exclusion principle obeyed by both systems. These expectation values capture only the population of each mode and are insensitive to the underlying commutation or anticommutation relations.

Consequently, at second order, the dynamics account only for direct, single-excitation exchange events between the system and the bath %i.e., processes generated solely by two-point bath correlation functions. 
Interference pathways -- such as those arising from higher-order processes involving multiple bath excitations or exchange loops -- are entirely absent. As a result, the evolution is insensitive to operator ordering, and the reduced dynamics of the system are indistinguishable between the two models. This agreement, however, is limited to second order; as we will see, differences between the fermionic and spin formulations begin to emerge at higher orders.\\

\subsubsection{Fourth-Order Master Equation}
\label{sec:fourth_order_ME}

At higher orders in the perturbative expansion, the structure of the reduced dynamics becomes qualitatively richer due to multiple excitation-exchange cycles\footnote{By ``excitation-exchange cycle'' we mean one emission/absorption event captured by a two-point bath correlator; at fourth order there are two such cycles, encoded in four-time correlators. No field-theoretic loop diagrams are implied.} involving repeated exchanges between the system and the bath. At fourth order, the system can interact with the environment in \emph{two} such exchange cycles before the bath is traced out, giving rise to terms involving multiple bath operators in four-time bath correlation functions. In such cases, the ordering of operators becomes relevant -- particularly when comparing fermionic and spin environments, whose operator algebras differ fundamentally.

A complete fourth-order master equation would require cataloging many contractions; that exercise, while straightforward, obscures the central point we wish to demonstrate. Instead, we focus on a single representative term, denoted \(\mathcal{K}_{\mathrm{fermion}}\) and \(\mathcal{K}_{\mathrm{spin}}\) for the two baths (defined and derived explicitly in Appendix~\ref{app:fourthorder}), which already captures all algebraic features distinguishing the two models. The key difference lies in the bath correlation functions that enter this representative term.

For the fermionic bath, the relevant correlation function is
\begin{widetext}
\begin{equation}\label{fermion_corr_main}
\mathcal{C}_{\text{fermion}}
\;=\;
\mathrm{Tr}_B\Bigl[
  c_{k_1}^\dagger\,
  c_{k_2}\,
  c_{k_3}\,
  c_{k_4}^\dagger
  \;\rho_B
\Bigr]
\;=\;
\delta_{k_1,k_2}\,\delta_{k_3,k_4}\,f_{k_1}(1-f_{k_3})
\;-\;
\delta_{k_1,k_3}\,\delta_{k_2,k_4}\,f_{k_1}(1-f_{k_2})
\;+\;
\text{(other nonvanishing terms)}.
\end{equation}
\end{widetext}
The additional terms include, in particular, the equal-mode case 
\(k_1=k_2=k_3=k_4\), which is generally nonzero.  
Because we assume a Gaussian fermionic bath, Wick’s theorem \cite{wick1950} applies, and all higher-order correlations reduce to products of two-point correlation functions.  
The crucial minus sign in the second term originates entirely from fermionic anticommutation and persists even if the additional nonvanishing terms are neglected.

For a spin bath, the corresponding correlation function reads
\begin{widetext}
\begin{equation}\label{spin_corr_main}
\mathcal{C}_{\text{spin}}
\;=\;
\mathrm{Tr}_B\Bigl[
  \sigma_{k_1}^\dagger\,
  \sigma_{k_2}\,
  \sigma_{k_3}\,
  \sigma_{k_4}^\dagger
  \;\rho_B
\Bigr]
\;=\;
\delta_{k_1,k_2}\,\delta_{k_3,k_4}\,f_{k_1}(1-f_{k_3})
\;+\;
\delta_{k_1,k_3}\,\delta_{k_2,k_4}\,f_{k_1}(1-f_{k_2})
\;+\;
\text{(other nonvanishing terms)}.
\end{equation}
\end{widetext}
Unlike the fermionic case, spin operators commute on different sites, so no minus sign appears. Moreover, spin baths are generally non-Gaussian, and their additional terms (including the equal-mode case) can encode genuine higher-order correlations absent in the Gaussian fermionic case.

The resulting representative fourth-order terms, \(\mathcal{K}_{\mathrm{fermion}}\) and \(\mathcal{K}_{\mathrm{spin}}\), derived explicitly in Appendix~\ref{app:fourthorder}, thus differ in two fundamental ways:  
(i) even when only the leading contributions are retained, fermionic anticommutation introduces relative minus signs in correlation functions that have no analogue in the spin case. This minus sign reflects the exchange statistics of indistinguishable fermions—a feature entirely absent in spin baths—and persists even if the additional nonvanishing terms are neglected;  
(ii) the structure of the additional nonvanishing terms is qualitatively different because the fermionic bath is Gaussian (Wick’s theorem applies), whereas spin baths are generally non-Gaussian and can have genuine higher-order correlations.

Consequently, although the second-order master equations for the two baths are identical, their fourth-order dynamics diverge significantly. The fermionic bath encodes antisymmetric exchange correlations through its operator algebra, whereas spin baths treat operators on different sites as commuting. This distinction becomes important when multi-excitation processes are relevant — particularly in strongly coupled or small baths — and motivates the use of exact-diagonalization methods discussed in the next subsection.

\subsection{Exact Diagonalization}
Our numerical analysis relies exclusively on exact diagonalization in the many--body basis \cite{jan1993}. In the fermionic model, each mode $j$ can be either occupied or unoccupied, resulting in a Fock-space dimension of $\binom{N_E + 1}{n_{\mathrm{exc}}}$ for a fixed total number $n_{\mathrm{exc}}$ of occupied modes. In the spin-model, each mode is similarly described by two states---occupied ($\ket{\uparrow}$) or unoccupied ($\ket{\downarrow}$)---and hence, for a given fixed number of occupied modes, the Hilbert space dimension remains the same as in the fermionic case. However, the fundamental differences between these representations emerge in the operator algebra rather than in the dimension of the Hilbert space itself.

Once the total Hamiltonian is represented in this many-body basis, it can be diagonalized to yield eigenstates and eigen--energies:
\begin{align}
H \ket{\Phi_\alpha} = E_\alpha \ket{\Phi_\alpha}.
\end{align}
The time evolution of an initial state $\ket{\psi(0)}$ is then computed as:
\begin{align}
\ket{\psi(t)} &= e^{-\mathrm{i} H t} \ket{\psi(0)} \notag \\
&= \sum_\alpha e^{-\mathrm{i} E_\alpha t}  \ket{\Phi_\alpha} \braket{\Phi_\alpha}{\psi(0)}.
\end{align}
From $\ket{\psi(t)}$, any observable of interest—such as the system population $\langle O_0^\dagger O_0 \rangle$—is readily computed. 

It is important to note, however, that the computational cost of exact diagonalization depends on the the size of the Hilbert space, which scales combinatorially. The dimension is maximal near half-filling, where it reaches \(\binom{N_E + 1}{(N_E + 1)/2}\) and scales exponentially with \(N_E\) (up to a sub-exponential prefactor). In contrast, configurations with very low or very high occupied numbers of bath modes correspond to significantly smaller subspaces. Accordingly, we restrict our analysis to small environments, where exact diagonalization remains computationally feasible, enabling direct numerical comparison between the fermionic and spin representations.

%%%%%%%%%%%%%%%%%%%%%%%%%%%%%%%%%RESULTS AND DISCUSSION%%%%%%%%%%%%%%%%%%%%%%%%%%%%%%%
\section{Numerical Results}
\label{sec_results_disc}
We begin by comparing the two models for a small bath, where differences are expected to be most pronounced. We consider $N_E=3$ bath modes with $n_{\mathrm{exc}}=2$ initially occupied, the system initially empty, and a uniform coupling $V_j=V=1$. 
The bath bandwidth is $\omega_{\mathrm{BW}}=2$, with equally spaced mode frequencies $\omega_1=-1$, $\omega_2=0$, and $\omega_3=1$. To make the algebraic origin of the differences transparent, we write the Hamiltonian in matrix form. Denoting the vacuum by $\ket{0}$, we label the six basis states $|0),\ldots,|5)$ as
\begin{align}
  |0) &= c_1^\dagger\,c_0^\dagger\,\ket{0},&
  |1) &= c_2^\dagger\,c_0^\dagger\,\ket{0},&
  |2) &= c_3^\dagger\,c_0^\dagger\,\ket{0}, \notag\\
  |3) &= c_2^\dagger\,c_1^\dagger\,\ket{0},&
  |4) &= c_3^\dagger\,c_1^\dagger\,\ket{0},&
  |5) &= c_3^\dagger\,c_2^\dagger\,\ket{0}.
\end{align}
Equivalently, we can define the same six states using spin operators. The action of the Hamiltonian, for example on $|0)$, is
\begin{widetext}
\begin{equation}
H|0) \;=\; H\,c_1^\dagger c_0^\dagger \ket{0}
= (\omega_0 + \omega_1)\,|0)
\;+\; V_2 \underbrace{c_2^\dagger c_0 c_1^\dagger c_0^\dagger \ket{0}}_{=-\,|3)}
\;+\; V_3 \underbrace{c_3^\dagger c_0 c_1^\dagger c_0^\dagger \ket{0}}_{=-\,|4)},
\end{equation}
\end{widetext}
where the minus signs are replaced by plus signs for spins because operators on different sites commute.

Collecting all matrix elements in this basis yields the following statistics-aware Hamiltonian matrix. Entries whose sign depends on operator reordering are written with $\mp$; take the upper $( - )$ sign for fermions and the lower $( + )$ sign for spins:
\begin{align}
&H \;=\; \notag\\
&\begin{bmatrix}
\omega_0 + \omega_1 & 0 & 0 & \mp V_2 & \mp V_3 & 0 \\
0 & \omega_0 + \omega_2 & 0 & \;\; V_1 & 0 & \mp V_3 \\
0 & 0 & \omega_0 + \omega_3 & 0 & \;\; V_1 & \;\; V_2 \\
\mp V_2 & \;\; V_1 & 0 & \omega_1 + \omega_2 & 0 & 0 \\
\mp V_3 & 0 & \;\; V_1 & 0 & \omega_1 + \omega_3 & 0 \\
0 & \mp V_3 & \;\; V_2 & 0 & 0 & \omega_2 + \omega_3
\end{bmatrix}.
\end{align}
This compact form makes the algebraic origin of the different dynamics explicit: the two models differ only by these statistics-dependent signs, so their time evolutions already diverge at the level of this finite matrix.

\begin{figure}[!h]
    \centering
    \includegraphics[width=\columnwidth]{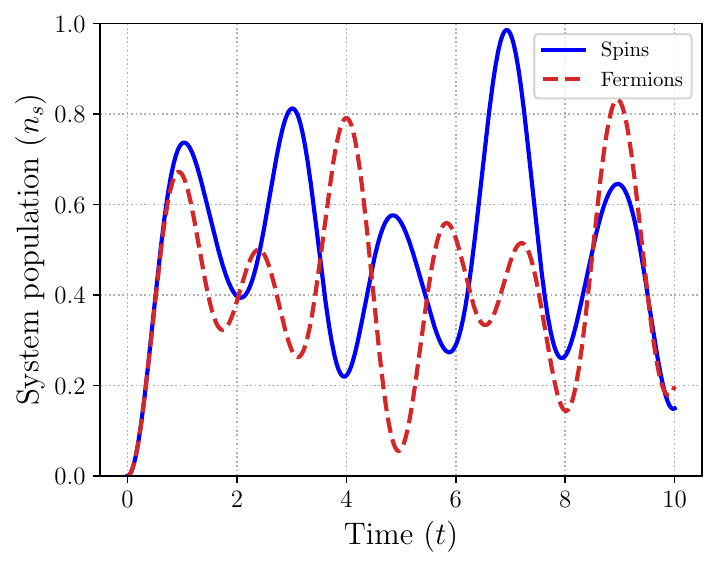}
    \caption{Population dynamics from exact diagonalization for $N_E=3$, $n_{\mathrm{exc}}=2$, uniform coupling $V=1$, and $\omega_{\mathrm{BW}}=2$ with mode frequencies $\{-1,0,1\}$. Fermionic results (red dashed) vs.\ spins (blue).}
    \label{fig:population_plot}
\end{figure}

Figure~\ref{fig:population_plot} shows the resulting system population $n_S(t)$ from exact diagonalization for the two models. We can see that even for this minimal bath, the two models display distinct oscillation and relaxation patterns.

In this model, if $n_{\mathrm{exc}}=1$ or $n_{\mathrm{exc}}=N_E$, then every nonzero off-diagonal element then only transfers that single excitation between the system and one bath mode. No process ever exchanges two indistinguishable bath particles, so no anticommutation sign arises. Consequently, the fermionic and spin Hamiltonians are identical in these cases, and the dynamics coincide.

%It is also worth noting that this is a minimal model, and if only one bath mode is occupied, or if all bath modes are occupied except one, there is effectively just a single excitation (or a single hole) that can move. Every nonzero off-diagonal element then only transfers that single excitation between the system and one bath mode. No process ever exchanges two indistinguishable bath particles, so no anticommutation sign arises. Consequently, the fermionic and spin Hamiltonians are identical in these cases, and the dynamics coincide.

We now turn to parameter choices relevant for larger environments. For an (effectively) infinite bath with equidistantly spaced modes, all coupled with the same constant \(V_k \equiv V\), Born–Markov master equations predict a golden-rule rate \cite{breuer2002theory,rivas2012open,Carmichael2008SMQO2}
\begin{align}
    \gamma \;=\; 2\pi\,|V|^{2}\,\rho(\omega_0),
\end{align}
where \(\rho(\omega_0)\) is the mode density at the system frequency. In our discrete bath with \(N_E\) modes uniformly spanning a bandwidth \(\omega_{\mathrm{BW}}\), \(\rho(\omega_0)=N_E/\omega_{\mathrm{BW}}\). We therefore parameterize the coupling by
\begin{align}\label{coupling-equ}
    V \;=\; \sqrt{\frac{\gamma}{2\pi\,\rho(\omega_0)}} \;=\; \sqrt{\frac{\gamma\,\omega_{\mathrm{BW}}}{2\pi\,N_E}}\,.
\end{align}
In the following results we set  $\omega_{\mathrm{BW}} = 4\gamma$, with the bath frequencies uniformly spaced within $\omega_{\mathrm{BW}}$.

\begin{figure}[!h]                                                          \includegraphics[width=\columnwidth]{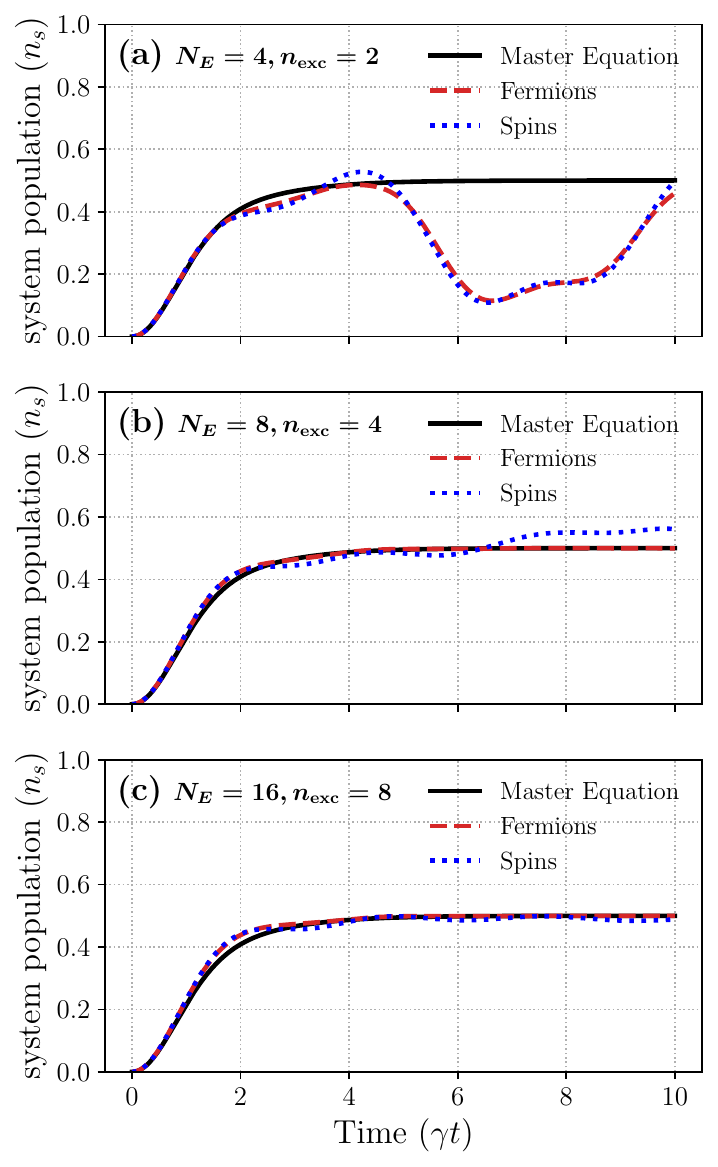}
    \caption{System population $n_S(t)$ with bandwidth $\omega_{\mathrm{BW}}=4\gamma$. Black solid: second-order master equation (ME2) red dashed: fermionic bath; blue dotted: spin bath. As $N_E$ increases, the fermionic and spin results converge and approach the ME2 prediction, reflecting the Markovian limit.}
    \label{fig:densitycoupling}
\end{figure}

Representative population dynamics for $N_E=4, 8, 16$ are shown in Fig.~\ref{fig:densitycoupling}. 
The system is initially unoccupied, and the bath is always half-filled.

The master‐equation results provide an analytic reference for the observed convergence. The black solid curves in Fig.~\ref{fig:densitycoupling} correspond to the second-order master equation (ME2) \cite{breuer2002theory, nakajima1958quantum, zwanzig1960ensemble} derived in Sec.~II, which, to second order in the system–bath coupling, is identical for fermions and spins. 
For small baths [panel (a)], both fermionic and spin results deviate noticeably from the ME2 prediction, reflecting the importance of higher-order exchange processes neglected by the master equation. 
These deviations are strongly influenced by interference pathways between multiple bath excitations. 
As the bath size increases [panels (b) and (c)], both exact diagonalization results approach the ME2 prediction, confirming that higher-order corrections become negligible when $N_E$ is large and the per-mode coupling decreases as $1/\sqrt{N_E}$. 
By $N_E=16$, the agreement is nearly quantitative, indicating that the dynamics can be fully captured by second-order processes in this effectively continuous-bath limit.

This convergence can also be interpreted in terms of bath statistics: the fermionic bath is Gaussian by construction, with all higher-order correlations reducible to products of two-point functions via Wick’s theorem. 
The spin bath, in contrast, is intrinsically non-Gaussian at small $N_E$, where additional higher-order correlations appear \cite{prokof2000theory, yang2008quantum, yang2009quantum}. 
However, as $N_E$ grows, the effective contribution of these higher-order spin correlations diminishes, and the bath approaches an effectively Gaussian description \cite{forsythe1999dissipative}. 
This behaviour is reminiscent of a central-limit-theorem-like tendency for the spin bath: while not proven explicitly here, the numerical convergence strongly suggests that the collective effect of many weakly coupled spin modes tends toward Gaussian statistics, explaining why both baths eventually yield identical Markovian dynamics and agree with the ME2 predictions.

\begin{figure}[!h]
    \centering
    \includegraphics[width=\columnwidth]{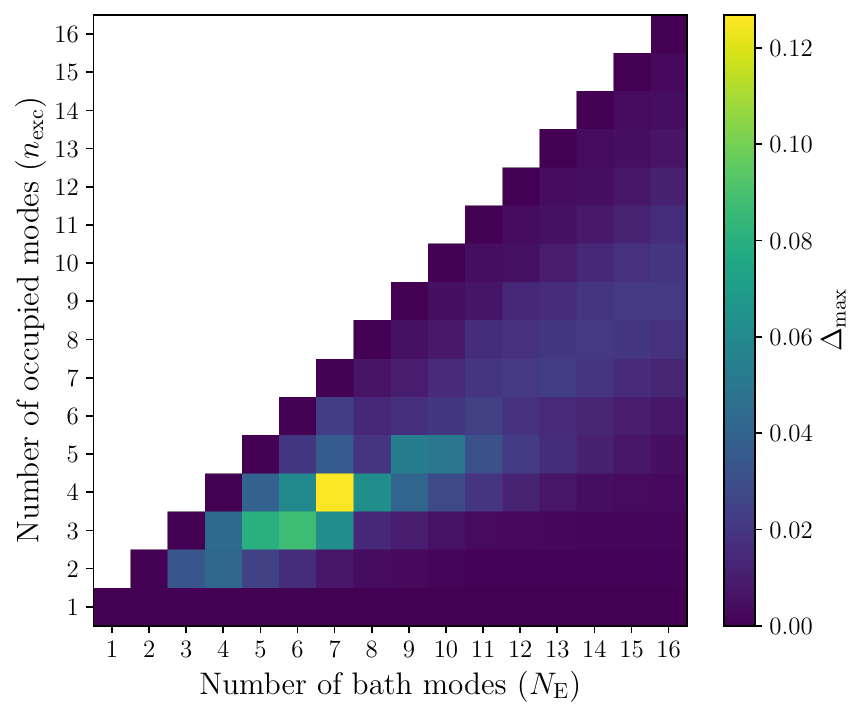}
    \caption{Maximum population difference $\Delta_{\mathrm{max}}$ between fermions and spins  with $\omega_{\mathrm{BW}}=4\gamma$, plotted as a function of $N_E$ (x-axis) and the number of occupied bath modes $n_{\mathrm{exc}}$ (y-axis). Differences peak at moderate bath sizes and near half filling, while vanishing for very small or large $N_E$.}
    \label{fig:heatmap}
\end{figure}

To systematically quantify the discrepancy between the two models, we define
\begin{align}
    \Delta_{\mathrm{max}} \;=\; \max_{\gamma t \in \{0,10\}} \Bigl|\, n_S^{(\mathrm{fermion})}(t) - n_S^{(\mathrm{spin})}(t) \Bigr|,
\end{align}
the maximum absolute difference in system population over time. 
Figure~\ref{fig:heatmap} shows a heatmap of $\Delta_{\mathrm{max}}$ plotted against bath size $N_E$ and number of initially occupied bath modes $n_{\mathrm{exc}}$. Note that the absolute magnitudes here are smaller than in Fig.~\ref{fig:population_plot}. This is because we now use a coupling defined in equ. (\ref{coupling-equ}), which is relatively smaller than the unscaled coupling $V=1$ used in Fig.~\ref{fig:population_plot}.

The largest discrepancies occur for moderate bath sizes ($4 \lesssim N_E \lesssim 8$) and near half-filling. 
In this regime, multiple bath modes participate in multi–loop exchange processes, with contributions that are strongly affected by fermionic antisymmetric operator ordering.
For very small or very large bath sizes, $\Delta_{\mathrm{max}}$ decreases: in small baths, only a limited number of exchange pathways exist, while in large baths the weak per-mode coupling suppresses higher-order contributions, leaving only second–order processes. 
When the bath is nearly empty or almost completely filled, $\Delta_{\mathrm{max}}$ is also small, since the number of available exchange pathways is minimal, leading to identical dynamics for fermions and spins.

\section{Conclusions}
\label{sec_conc}
We compared a minimal resonant-level model coupled to either a fermionic or a spin bath using exact diagonalization and an analytically tractable second-order master equation (ME2).  Importantly, the reduced dynamics are already sensitive to operator ordering, despite the model being strictly quadratic.  The two descriptions agree at second order—where only two-point correlators enter—but differ once higher-order processes contribute.  In small and moderately sized baths, especially near half-filling, antisymmetric fermionic operator algebra introduces exchange minus signs that suppress certain multi-excitation pathways, while commuting spin operators make those pathways add constructively, yielding different dynamics. 

As the bath size increases while the per-mode coupling decreases as $V \propto N_E^{-1/2}$ for a fixed Markovian rate $\gamma$, higher-order corrections become negligible and both models converge to the ME2 prediction. Consistent with this, we observe that many weakly coupled spin modes collectively mimic Gaussian fermionic statistics in the effectively Markovian regime, even though the baths differ microscopically.

Practically, spin-bath simulations are reliable for large, weakly coupled environments, but for smaller or more strongly coupled baths—particularly near half-filling—a fully fermionic treatment is required to capture exchange-statistics–driven effects in the reduced dynamics.\\

\section*{Acknowledgments}
M.C. acknowledges support by the Return Program of the State of North Rhine-Westphalia. M.Z acknowledges the James and Enid Nicol Trust for a PhD scholarship.

\appendix
%%%%%%%%%%%%%%%%%%%%%%%%%%%%%%%%%%%%%%%%%%%%%%%%%%%%%%%%%%%%%%%%%%%%%%%%%%%%
% Appendix B: 2nd order ME
%%%%%%%%%%%%%%%%%%%%%%%%%%%%%%%%%%%%%%%%%%%%%%%%%%%%%%%%%%%%%%%%%%%%%%%%%%%%

\section{Second-Order Master Equation}
\label{app:secorder_equiv}
Consider the total Hamiltonian defined in Sec.~\ref{sec_theory} as
\(H = H_S + H_E\), where \(H_E\) bundled together the free bath and the
system–bath coupling. For the master–equation derivation it is convenient
to split this into a free part and an interaction:
\begin{equation}
    H \;=\; H_S + H_B + H_{SB},
\end{equation}
where the system Hamiltonian is
\begin{equation}
    H_S \;=\; \omega_0\, O_0^\dagger O_0,
\end{equation}
the bath Hamiltonian is
\begin{equation}
    H_B \;=\; \sum_k \omega_k\, O_k^\dagger\,O_k,
\end{equation}
and the system-bath interaction is given by
\begin{equation}
    H_{SB} \;=\; \sum_k \Bigl( V_k\,O_0^\dagger\,O_k + V_k^*\,O_k^\dagger\,O_0 \Bigr).
\end{equation}
Here, \(O_0\) may represent either a fermionic annihilation operator \(c_0\) or a spin--like lowering operator \(\sigma_0\). Likewise, \(O_k\) is the corresponding bath operator (\(c_k\) or \(\sigma_k\)), with \(\hbar=1\) throughout.

Moving to the interaction picture, the free evolution of the system and bath operators is
\begin{align}
    O_0(t) &= e^{-\,i\,\omega_0\,t}\,O_0,\\
    O_k(t) &= e^{-\,i\,\omega_k\,t}\,O_k,
\end{align}
and thus the interaction Hamiltonian becomes
\begin{equation}
    \tilde{H}_{SB}(t) 
    \;=\; 
    \sum_k \Bigl[
       V_k\,e^{\,i(\omega_0 - \omega_k)\,t}\,O_0^\dagger\,O_k
       \;+\; \mathrm{h.c.}
    \Bigr].
\end{equation}
Assuming an initially factorized system-bath state, the second-order contribution to the master equation for the reduced density matrix \(\rho_S(t)\) is
\begin{equation}
    \frac{d\rho_S(t)}{dt}
    \;=\;
    -\int_0^t dt'\,\mathrm{Tr}_B\,\Bigl[
      \tilde{H}_{SB}(t),\;
      \bigl[\tilde{H}_{SB}(t'),\;\rho_S(t') \otimes \rho_B\bigr]
    \Bigr].
\end{equation}
Expanding the nested commutators and discarding vanishing terms, one obtains
\begin{widetext}
    \begin{align}
   \frac{d\rho_S(t)}{dt}&
   \;=\;
   -\sum_k |V_k|^2 \int_0^t dt' \nonumber\\
   \times \Bigl[&
       e^{\,i\,\omega_k\,(t-t')}\,\langle O_k^\dagger O_k\rangle\,
       e^{-\,i\,\omega_0\,(t-t')}
       \,\bigl[\,O_0,\;O_0^\dagger\,\rho_S(t')\bigr] +
       e^{\,i\,\omega_k\,(t-t')}\,\langle O_k O_k^\dagger\rangle\,
       e^{-\,i\,\omega_0\,(t-t')}
       \,\bigl[\,\rho_S(t')\,O_0^\dagger,\;O_0\bigr]
   \notag \\
   &+\;
       e^{-\,i\,\omega_k\,(t-t')}\,\langle O_k O_k^\dagger\rangle\,
       e^{\,i\,\omega_0\,(t-t')}
       \,\bigl[\,O_0^\dagger,\;O_0\rho_S(t)\bigr]
+
       e^{-\,i\,\omega_k\,(t-t')}\,\langle O_k^\dagger O_k\rangle\,
       e^{\,i\,\omega_0\,(t-t')}
       \,\bigl[\,\rho_S(t)\,O_0,\;O_0^\dagger\bigr]
   \Bigr].
\end{align}
\end{widetext}

We assume the bath is in thermal equilibrium, governed by a Fermi--Dirac distribution:
\[
   f(\omega_k) 
   \;=\;
   \frac{1}{e^{\beta(\omega_k-\mu)} + 1}.
\]
Then, the correlation functions are
\begin{align}
    \langle O_k\,O_j^\dagger\rangle 
    &= 
    \delta_{k j}\,\bigl[\,1 - f(\omega_k)\bigr],\\
    \langle O_k^\dagger\,O_j\rangle 
    &= 
    \delta_{k j}\,f(\omega_k).
\end{align}
Defining
\begin{align}
    C_1(\tau) &= \sum_k |V_k|^2 \,e^{-\,i(\omega_0-\omega_k)\,\tau}\,f(\omega_k),\\
    C_2(\tau) &= \sum_k |V_k|^2 \,e^{-\,i(\omega_0-\omega_k)\,\tau}\,\bigl[1 - f(\omega_k)\bigr],
\end{align}
and letting \(\tau = t - t'\), one can rewrite the master equation compactly as
\begin{align}
    \frac{d\rho_S(t)}{dt} 
    \;=\; 
    -\int_0^t& dt' \,\Bigl(
       C_1(\tau)\,\bigl[\,O_0,\;O_0^\dagger\,\rho_S(t')\bigr] \nonumber\\
       &\;+\;C_2(\tau)\,\bigl[\,\rho_S(t')\,O_0^\dagger,\;O_0\bigr]
       \;+\; \mathrm{h.c.}    \Bigr).
\end{align}
Crucially, this second-order master equation depends only on bath correlation functions, which are entirely specified by the bath statistics and do not hinge on the (anti)commuting nature of \(O_k\). As a result, the same expression holds whether \(O_0\) and \(O_k\) are fermionic annihilation operators or spin--like lowering operators.

%%%%%%%%%%%%%%%%%%%%%%%%%%%%%%%%%%%%%%%%%%%%%%%%%%%%%%%%%%%%%%%%%%%%%%%%%%%%
% Appendix C: 4th order ME
%%%%%%%%%%%%%%%%%%%%%%%%%%%%%%%%%%%%%%%%%%%%%%%%%%%%%%%%%%%%%%%%%%%%%%%%%%%%
\section{Fourth-Order Master Equation}
\label{app:fourthorder}

In this appendix, we carefully look at a fourth-order contribution to the perturbative master equation, explicitly addressing the crucial aspects of operator ordering in the bath correlation functions. The starting point is the total Hamiltonian in the interaction picture:
\begin{equation}\label{eq:HSB}
  \tilde{H}_{SB}(t) 
  \;=\; 
  \sum_{k} \Bigl[
    V_{k}\, O^\dagger O_{k}\, e^{\,i\Delta_{k} t}
    \;+\;
    V_{k}^*\, O_{k}^\dagger O\, e^{-\,i\Delta_{k} t}
  \Bigr],
\end{equation}
with
\[
\Delta_k \;=\; \omega_0 - \omega_k,
\]
where \(O\) and \(O^\dagger\) denote system operators, while \(O_k\) and \(O_k^\dagger\) act on the bath. Our goal is to analyze the perturbative expansion of the master equation to fourth order in \(H_{SB}\), focusing on a representative terms that yield nonzero contributions under the bath trace. We retain an abstract operator notation for \(O\) and \(O_k\) until we later specify whether the bath operators are fermionic or spin.

The reduced dynamics of the system can be obtained from the Liouville--von Neumann equation via a perturbative expansion. Specifically, the time evolution of the system density matrix \(\rho_S(t)\) may be written as
\begin{equation}\label{ME_expansion}
    \frac{d\rho_S(t)}{dt}
    \;=\;
    \sum_{n=1}^\infty \mathcal{D}^{(n)}(t),
\end{equation}
where \(\mathcal{D}^{(n)}(t)\) denotes the \(n\)th--order contribution in the system--bath coupling. At fourth order, the contribution takes the schematic form
\begin{widetext}
\begin{equation}\label{D4}
\begin{split}
    \mathcal{D}^{(4)}(t) \;=\;
    (-i)^4 \!
    \int_0^t dt_1 
    \!\int_0^{t_1} dt_2 
    \!\int_0^{t_2} dt_3 
    \!\int_0^{t_3} dt_4\; 
    \mathrm{Tr}_B \Bigl[
      \tilde{H}_{SB}(t_1),\;
      \Bigl[
        \tilde{H}_{SB}(t_2),\;
        \Bigl[
          \tilde{H}_{SB}(t_3),\;
          \Bigl[
            \tilde{H}_{SB}(t_4),\;
            \rho_S(t_4) \otimes \rho_B
          \Bigr]
        \Bigr]
      \Bigr]
    \Bigr].
\end{split}
\end{equation}
\end{widetext}
The nested time integrals explicitly enforce the causal ordering $
t_1 > t_2 > t_3 > t_4,$
which is equivalent to the action of the standard time--ordering operator \(\mathcal{T}\) in the Dyson series. Thus, \(\mathcal{T}\) does not appear explicitly in Eq.~\eqref{D4}. Although the nested commutators generate 16 distinct terms, we now focus on one representative nonvanishing term. In particular, we consider
\begin{equation}\label{repTerm}
\mathrm{Tr}_B \Bigl[
  \tilde{H}_{SB}(t_1)\,\tilde{H}_{SB}(t_2)\,\tilde{H}_{SB}(t_3)\,\tilde{H}_{SB}(t_4)
  \;\rho_S(t)\,\otimes\,\rho_B
\Bigr],
\end{equation}
which describes a sequence of four interactions between the system and the bath. Substituting Eq.~\eqref{eq:HSB} into this product and simplifying yields 16 terms, only some of which survive under the bath trace.

At a given time \(t_i\), we write
\begin{align}\label{HSB-ki}
\tilde{H}_{SB}(t_i) 
&=\;
\sum_{k_i} \Bigl[
  V_{k_i}\, O^\dagger O_{k_i}\, e^{\,i\,\Delta_{k_i}\,t_i}
  \;+\;
  V_{k_i}^*\, O_{k_i}^\dagger O\, e^{-\,i\,\Delta_{k_i}\,t_i}
\Bigr],
\end{align}
with $
\Delta_{k_i} = \omega_0 - \omega_{k_i}
$.
Substituting Eq.~\eqref{HSB-ki} into the nested commutators yields a product of four such interaction Hamiltonians. After simplifying, one encounters another 16 terms, and we focus on a representative nonvanishing term:
\begin{widetext}
\begin{equation}\label{repTerm2}
\begin{split}
\mathcal{K} (t_1, t_2, t_3, t_4) 
&=\;
\sum_{k_1,k_2,k_3,k_4}
\Bigl[
  V_{k_1}\,V_{k_2}^*\,V_{k_3}\,V_{k_4}^*\,
  e^{\,i\,(-\Delta_{k_1}\,t_1 + \Delta_{k_2}\,t_2 + \Delta_{k_3}\,t_3 - \Delta_{k_4}\,t_4)}
  O^\dagger_{k_1} \, O\,
  O^\dagger \,O_{k_2}\,
  O^\dagger\,O_{k_3}\,
  O_{k_4}^\dagger\,O
  \;\rho_S
  \;\otimes\;\rho_B
\Bigr].
\end{split}
\end{equation}
%\end{widetext}
To isolate the essential operator structure pertaining to the bath, we may rewrite \(\mathcal{K}\) in the form
%\begin{widetext}
\begin{equation} \label{T_with_Cbath}
\mathcal{K} (t_1, t_2, t_3, t_4) 
\;=\;
\sum_{k_1,k_2,k_3,k_4}
\Bigl[
V_{k_1}\,V_{k_2}^*\,
V_{k_3}\,V_{k_4}^*\,
e^{\,i\,(-\Delta_{k_1}\,t_1 + \Delta_{k_2}\,t_2 + \Delta_{k_3}\,t_3 - \Delta_{k_4}\,t_4)}
\;
\bigl(O \, O^\dagger O^\dagger\, O\bigr)\,\rho_S
\;\times\;
\mathcal{C}_{\mathrm{bath}}\Bigr],
\end{equation}
\end{widetext}
where the bath--dependent part is
\begin{equation}\label{Cbath}
\mathcal{C}_{\mathrm{bath}}
\;=\;
\mathrm{Tr}_B\Bigl[
  O_{k_1}^\dagger\,
  O_{k_2}\,
  O_{k_3}\,
  O_{k_4}^\dagger
  \;\rho_B
\Bigr].
\end{equation}
From this point onward, we focus on the evaluation of \(\mathcal{C}_{\mathrm{bath}}\), while understanding that the time--dependent phases from Eq.~\eqref{repTerm2} factor out.

\medskip

To determine which index combinations \(\{k_1,k_2,k_3,k_4\}\) yield a nonzero trace, we assume a product-state bath density matrix,
\begin{equation}\label{rhoB_product}
\rho_B 
\;=\;
\bigotimes_{k=1}^N\,\rho_B^{(k)},
\qquad
\rho_B^{(k)} 
\;=\;
(1-f_k)\,\ket{0}_k\bra{0}
\;+\;
f_k\,\ket{1}_k\bra{1},
\end{equation}
with $f_k \equiv f(\omega_k)$. Therefore, only terms with an equal number of creation and annihilation operators on each mode contribute to the trace.

For a fermionic bath, where \(\{c_k, c_{k'}^\dagger\} = \delta_{kk'}\), we evaluate representative contributions explicitly. Three illustrative cases are:

\paragraph*{Case 1: \(k_1 = k_2\), \(k_3 = k_4\), and \(k_1 \neq k_3\).}
\begin{align}\label{fermion_case1}
\mathcal{C}^{(1)}_\text{fermion}
&=\; \delta_{k_1,k_2}\,\delta_{k_3,k_4} \,
\mathrm{Tr}\Bigl[c_{k_1}^\dagger\,c_{k_1}\,\rho_B^{(k_1)}\Bigr]
\;
\mathrm{Tr}\Bigl[c_{k_3}\,c_{k_3}^\dagger\,\rho_B^{(k_3)}\Bigr]
\nonumber\\
&=\;
\delta_{k_1,k_2}\,\delta_{k_3,k_4}\,f_{k_1}\,\bigl(1-f_{k_3}\bigr).
\end{align}

\paragraph*{Case 2: \(k_1 = k_3\), \(k_2 = k_4\), and \(k_1 \neq k_2\).}
\begin{align}\label{fermion_case2}
\mathcal{C}^{(2)}_\text{fermion}
&=\; -\,\delta_{k_1,k_3}\,\delta_{k_2,k_4}\,
\mathrm{Tr}\Bigl[c_{k_1}^\dagger\,c_{k_1}\,\rho_B^{(k_1)}\Bigr]
\;
\mathrm{Tr}\Bigl[c_{k_2}\,c_{k_2}^\dagger\,\rho_B^{(k_2)}\Bigr]
\nonumber\\
&=\;
-\delta_{k_1,k_3}\,\delta_{k_2,k_4}\,f_{k_1}\,\bigl(1-f_{k_2}\bigr).
\end{align}

\paragraph*{Case 3: \(k_1 = k_4\), \(k_2 = k_3\), and \(k_1 \neq k_2\).}
\begin{align}\label{fermion_case3}
\mathcal{C}^{(3)}_\text{fermion}
&=\;
\mathrm{Tr}\Bigl[c_{k_1}^\dagger\,c_{k_1}^\dagger\,\rho_B^{(k_1)}\Bigr]
\;
\mathrm{Tr}\Bigl[c_{k_2}\,c_{k_2} \,\rho_B^{(k_2)}\Bigr]
\;=\;0.
\end{align}

Summing all contributions, we write
\begin{widetext}
\begin{equation}\label{fermion_total_final}
\begin{split}
\mathcal{C}_\text{fermion}
&=\;
\mathcal{C}^{(1)}_\text{fermion}
\;+\;
\mathcal{C}^{(2)}_\text{fermion}
\;+\;
\mathcal{C}^{(3)}_\text{fermion}
\;+\;
\text{(other nonvanishing terms)}
\\
&=\;
\delta_{k_1,k_2}\,\delta_{k_3,k_4}\,f_{k_1}(1-f_{k_3})
\;-\;
\delta_{k_1,k_3}\,\delta_{k_2,k_4}\,f_{k_1}(1-f_{k_2})
\;+\;
\text{(other nonvanishing terms)},
\end{split}
\end{equation}
\end{widetext}
where the additional terms include, in particular, the case \(k_1=k_2=k_3=k_4\), which is generally nonzero for our case.

For spin baths, with \([\sigma_k,\sigma_{k'}^\dagger]=\delta_{kk'}\), the same index patterns contribute, but no minus sign appears:
\begin{widetext}
\begin{equation}\label{spin_total_final}
\mathcal{C}_\text{spin}
=
\delta_{k_1,k_2}\,\delta_{k_3,k_4}\,f_{k_1}(1-f_{k_3})
\;+\;
\delta_{k_1,k_3}\,\delta_{k_2,k_4}\,f_{k_1}(1-f_{k_2})
\;+\;
\text{(other nonvanishing terms)}.
\end{equation}
\end{widetext}
The additional nonvanishing terms, such as the case \(k_1=k_2=k_3=k_4\), are not written explicitly here. 
Their detailed structure differs between fermionic and spin baths because the fermionic bath is assumed Gaussian, whereas spin baths are generally non-Gaussian. 

However, even if these terms are neglected or vanish, the fermionic and spin correlations remain distinct: the fermionic case carries relative minus signs from anticommutation (as in Case~2), whereas spin operators commute. 
Thus, the representative fourth-order terms in the master equation differ for the two types of baths, even when their spectral densities and occupations are identical.
    
\section{Single--Body Diagonalization for Fermions vs.\ Spin Models}
\label{app:singlebody}

A key simplification of the fermionic formulation is that the Hamiltonian is quadratic in creation and annihilation operators,
\[
    H = \sum_{i,j} A_{ij}\, c_i^\dagger c_j,
\]
with
\[
    \{c_i, c_j^\dagger\} = \delta_{ij}, 
    \quad 
    \{c_i, c_j\} = 0.
\]
The single--particle problem defined by the Hermitian matrix \(A\) can be diagonalized via a unitary transformation
\[
    c_i = \sum_\alpha P_{i\alpha}\, d_\alpha,
    \quad
    c_i^\dagger = \sum_\alpha d_\alpha^\dagger\,P_{i\alpha}^*.
\]
The new operators obey
\[
    \{d_\alpha, d_\beta^\dagger\}
    = (P^\dagger P)_{\alpha\beta}
    = \delta_{\alpha\beta},
    \quad
    \{d_\alpha, d_\beta\} = 0,
\]
so the fermionic algebra is preserved.  
In the new modes \(d_\alpha\), the Hamiltonian becomes diagonal,
\[
    H = \sum_\alpha \lambda_\alpha\, d_\alpha^\dagger d_\alpha.
\]
Population dynamics and correlation functions can then be computed directly from this single--particle diagonalization without constructing the full many--body Hilbert space.

\vspace{0.5em}
For the spin model, the bath modes are described by ladder operators \(\{\sigma_i^-, \sigma_i^+\}\) obeying
\[
    \{\sigma_i^-, \sigma_i^+\} = 1,
    \quad
    [\sigma_i^\pm, \sigma_j^\pm] = 0 \quad (i \neq j).
\]
These relations mix commutation and anticommutation depending on whether operators act on the same or different sites.  
If one attempts a basis transformation
\[
    \sigma_i^- = \sum_\alpha P_{i\alpha} \, d_\alpha,
    \quad
    \sigma_i^+ = \sum_\alpha d_\alpha^\dagger \, P_{i\alpha}^*,
\]
the resulting mode operators satisfy
\[
    \{d_\alpha, d_\beta^\dagger\}
    = (P^\dagger P)_{\alpha\beta}
      + 2 \sum_{k \neq \ell} P_{k\alpha}^* P_{\ell\beta} \, \sigma_\ell^+ \sigma_k^-,
\]
where the second term is operator--valued.  
This means the transformed operators do \emph{not} obey simple canonical anticommutation relations, and the Hamiltonian cannot be reduced to a diagonal sum of commuting bilinear terms.  
As a result, the efficient single--particle diagonalization applicable to fermions does not carry over to the spin model, which must be treated by explicit many--body diagonalization or other approximation schemes.

\bibliography{apssamp}% Produces the bibliography via BibTeX.

\end{document}